# Cavitation nanopore in the dielectric fluid in the inhomogeneous, pulsed electric fields

Mikhail Pekker [1] and Mikhail N. Shneider [2*]

[1] *MMSolution, 6808 Walker Street, Philadelphia, PA, 19135*
[2] *Department of Mechanical and Aerospace Engineering, Princeton University, Princeton, NJ, 08544*

**Abstract**
*This paper discusses the nanopores emerging and developing in a liquid dielectric under the action of the ponderomotive electrostrictive forces in a nonuniform electric field. It is shown that the gradient of the electric field in the vicinity of the rupture (cavitation nanopore) substantially increases and determines whether the rupture grows or collapses. The cavitation rupture in the liquid (nanopore) tends to stretch along the lines of the original field. The mechanism of the breakdown associated with the generation of secondary ruptures in the vicinity of the poles of the nanopore is proposed. The estimations of the extension time for nanopore in water and oil (polar and nonpolar liquids, respectively) are presented. A new mechanism of nano- and subnanosecond breakdown in the insulating (transformer) oil that can be realized in the vicinity of water microdroplets in modern nanosecond high-voltage devices is considered.*

**I. Introduction.**

The problem of expansion (compression) of a gas bubble in water in the approximation of incompressible fluid without friction was considered by Rayleigh in the classic work [1]. The effect of viscosity on the dynamics of the bubble oscillation was studied both experimentally and theoretically by many authors (see, e.g., [2-5]). Statistical theory of the bubbles emerging at stretching negative pressure has been developed by Zel'dovich [6] and Fisher [7]. Effect of electric field on the form of a gas bubble in an electric field was explored in a number of papers (see for example [8-10]). The main result of these studies is that a spherical bubble is transformed into an ellipsoid, which is elongated along the electric field.

The discharge development in a dielectric liquid (water) at nanosecond pulse voltage applied to a sharp needle electrode remained until recently a mysterious phenomenon. That is because the high-voltage pulse time ~ 1-5 ns is not enough for the formation of vapor bubbles, where a regular "gas-discharge" electrical breakdown occurs, as in the case with microsecond voltage pulses [11-12]. A few ideas were suggested to explain this phenomenon (see e.g. [13-17]). However, they did not give a reasonable explanation of the nanosecond breakdown in the liquid.

In [18] there was proposed an electrostrictive mechanism of the nanosecond breakdown in dielectric fluid, and the corresponding experiment, which allows to check it. The idea of [18] was following: at the nanosecond pulse voltage applied to the needle electrode, a stretching tension occurs in the fluid which leads to disruption of its continuity, i.e. formation of cavitation nanopores in which an electron can gain the energy necessary for ionization. The main difference between fast (nanosecond) pulses from slow (microsecond) is that in the time of a few nanoseconds, the liquid due to inertia does not have time to displace to the electrode and to compensate the negative pressure [19]. The work [18] has stimulated numerous experimental and theoretical studies on nanosecond breakdown in liquid dielectrics [20-25].

Note that only the appearance of ruptures in the fluid is not sufficient for the development of breakdown, since at the radius of the nanopore $R \sim 1-2$ nm an electron therein cannot acquire energy necessary for the ionization of the water molecules, i.e. $2E_{in} \cdot R < 12.6 \, \text{eV}$ ($E_{in}$ is the field inside the pore). At the same time the pores expansion is suppressed by Laplace pressure $p_L = 2\sigma_{eff}/R$, which reaches tens of MPa at the pore radius $R \sim 1-5$ nm and, therefore, cannot be compensated by the saturated vapor pressure. For micro- and nanopores $\sigma_{eff} = k_\sigma \sigma$ is the effective surface tension, which takes into account the Tolman correction factor $k_\sigma(R)$ [26]. We assume for estimates, without loss of generality, $k_\sigma = 1$. As it was shown in [27], an electrostatic pressure arises in the cavitation pore, due to the distortion of its main electric field, which exceeds the Laplace pressure. In the paper [28], the estimate of the rate of the nanopores expansion is made, and it was shown that three characteristic regions are arising in the vicinity of a needle electrode. In the first region, where the electric field gradient is the greatest, the occurring cavitation nanopores have enough time to grow to a size at which an electron can gain enough energy for the excitation and ionization of the liquid molecules on the pore wall. In the second region, the electrostrictive negative pressure reaches values at which the cavitation development becomes possible, but the nanopores, appearing during the voltage pulse, do not have enough time to grow to the size at which the potential difference across their borders becomes sufficient for the ionization or excitation of water molecules. And, in the third one, the development of cavitation is impossible, since the spontaneously occurring nanovoids do not grow, because the value of the electrostrictive negative pressure is relatively small and cannot compete with the forces of surface tension.

---

[*] m.n.shneider@gmail.com

At present, qualitative changes occur in the high-voltage pulse technique: the steepness of the voltage pulse increases and the front duration decreases to nanoseconds or even to hundreds of picoseconds. Therefore, there is a great practical interest to the pulsed dielectric strength of high voltage insulation. In the high-voltage devices the transformer oil is commonly used as an insulating material.

This paper shows that the cavitation rupture in liquids (nanopore) tends to stretch in the direction of the external field. The mechanism of sub- and nanosecond breakdown, associated with the generation of secondary ruptures in the poles vicinity of the original nanopores, is proposed. The estimates of the expansion time of the nanopores in water and oil (in polar and nonpolar liquids, respectively) are presented. It is shown that it is possible the development of nano- and sub-nanosecond breakdown in oil in the vicinity of the poles of microdroplets of water, which significantly increases the requirements for purity of insulating oil used in nanosecond high-voltage devices.

**1. Estimation of the rate of the nanopores expansion.**
Let us assume a spherical nanocavity of the radius $R$ is formed in water as a result of thermal fluctuations. Neglecting with gas pressure and considering the discontinuity of the dielectric constant at the liquid-vacuum boundary, the pressure exerted by the electric field on the surface of the pore per unit area [29]:

$$P_n = -P_E \left(\frac{3\varepsilon}{1+2\varepsilon}\right)^2 \left(\left(1-\left(1-\frac{1}{\varepsilon^2}\right)\cos^2(\theta)\right) - \frac{1}{\alpha}\left(1-\frac{1}{\varepsilon}\right)\left(1-\left(1-\frac{1}{\varepsilon}\right)\cos^2(\theta)\right)\right) - p_h - \frac{2\sigma}{R} \quad (1)$$

Here, $P_E = -\frac{1}{2}\alpha\varepsilon_0\varepsilon E_0^2$, $E_0$ is unperturbed electric field at the pore location, $\varepsilon$ is the dielectric permittivity of the liquid, $\alpha \sim 1.3-1.5$ is the empirical factor for most of the studied polar dielectric liquids, including water, [12], and $\alpha = (\varepsilon-1)\cdot(\varepsilon+2)/(3\varepsilon)$ for nonpolar liquid [30]; $p_h$ is the hydrostatic pressure on the surface of the pore. We have taken into account in (1) that since the critical initial sizes of the cavitation pore is about a few nanometers, and the unperturbed electric field induced by the high voltage potential of the electrode is changing on a length scale of order of the electrode size (~ 10-100 microns, typical for experimental conditions in [20-25]), therefore, the external unperturbed electric field $E_0$ in the vicinity of the pore can be considered as homogeneous with a good accuracy $E_0 /( dE_0 / dr ) \gg R$, and the electric potential inside of the pore is $\varphi_{in} = -\frac{3\varepsilon}{\varepsilon_{in}+2\varepsilon} E_0 r \cos(\theta)$, $r \leq R$, and outside $\varphi_{out} = -E_0 r \cos(\theta)\left(1+\frac{(\varepsilon-\varepsilon_{in})}{2\varepsilon+\varepsilon_{in}}\frac{R^3}{r^3}\right)$, $r > R$ [30].

Since the pressure in the fluid associated with the ponderomotive force is $P_{pond} = -\frac{1}{2}\alpha\varepsilon_0\varepsilon E^2$, then, by calculating the electric field $\bar{E} = -\nabla\varphi$ outside spherical pores, the total pressure in the fluid outside the pores:

$$P_{total} = p + P_0 + P_p = p + P_E + P_E\left((3\cos^2(\theta)+1)\left(\frac{(\varepsilon-\varepsilon_{in})}{2\varepsilon+\varepsilon_{in}}\right)^2\frac{R^6}{r^6} - (5\cos^2(\theta)-1)\frac{R^3}{r^3}\frac{(\varepsilon-\varepsilon_{in})}{2\varepsilon+\varepsilon_{in}}\right) \quad (2)$$

Here $p$ is the hydrostatic pressure in the liquid; $P_0 = P_E, P_p$ are the electrostrictive pressures, which are related to the presence of the undisturbed electric field and its perturbation in the vicinity of the pore, correspondingly; $\varepsilon_{in} = 1$. At $r = R$, the pressure $p$ in (2) is equal to $p_h$ in (1). The ponderomotive pressures distributions in the vicinity of the pore for water, $\varepsilon = 80$ [31], and transformer oil, $\varepsilon = 2.3$ [32], normalized to the corresponding values $|P_E|$, are shown in Fig. 1.

As noted above, during the nanosecond voltage pulse on the needle electrode, the liquid does not have time to be shifted to the electrode, and to compensate the electrostrictive pressure of the external electric field (second term in (2)). However, the compensation of the electrostrictive stretching, related to the potential redistribution caused by the micropore itself, is much faster. From the linearized equations of the fluid motion $\rho\partial v/\partial t \sim P_E/R$ can be estimated the time of the fluid influx to the pore, $t_p \sim R\sqrt{\rho/|P_E|}$. For example, in water, for the pore with radius $R = 2$ nm the stretching forces exceed the Laplace pressure at $|P_E| \geq 250$ MPa ( $E_0 \geq 8\cdot 10^8$ V/m), $t_p \sim 4$ ps. Assuming that the characteristic time of the electrostrictive pore expansion is smaller than the time for the fluid influx, in formula (1) we can set $p_h = P_p(R,\theta)$. Fig. 2 shows the angular dependence of the normal pressure $P_n$ acting on the surface of the pores without taking into account the surface tension forces for water and transformer oil, correspondingly. It is seen that the pore in water as well as in transformer oil is stretched along the electric field, since the pressure at its poles is greater than the pressure at the equator.

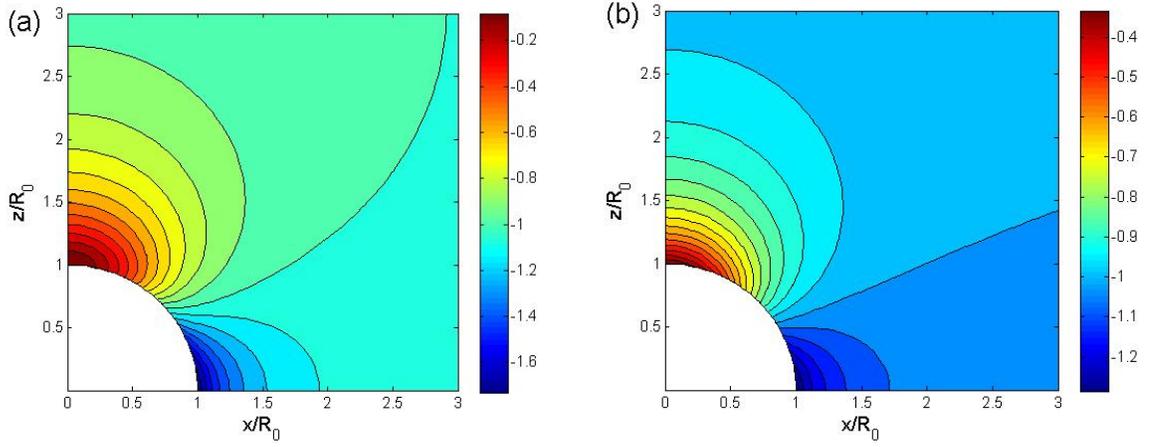

**Fig. 1**. Two-dimensional distribution of the electrostrictive pressure outside of the pore, normalized to $|P_E|$; (a) – in water, (b) – in transformer oil.

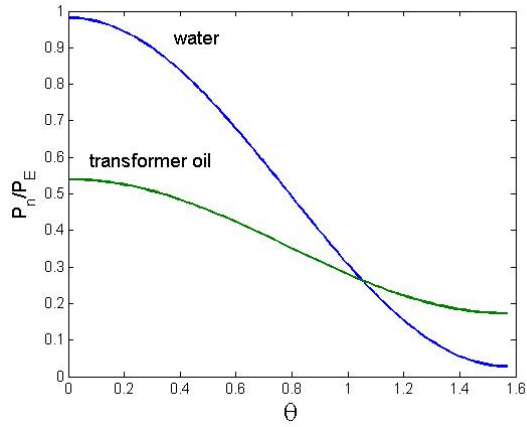

**Fig. 2.** The angular dependence of $P_n$ without taking into account the surface tension.

In the approximation of a spherical pore, from the equations (1) and (2) and after averaging over the angle of all the forces acting on the surface, we obtain the equation for the radius of the expanding pores [28]:

$$\frac{d}{dt}\left(R^3\left(\frac{dR}{dt}\right)^2\right) =$$

$$\frac{2R^2 P_E}{\rho}\frac{dR}{dt}\left(\left(\frac{3\varepsilon}{1+2\varepsilon}\right)^2\left(\left(\frac{1}{3}\frac{1}{\varepsilon^2}+\frac{2}{3}\right)-\frac{(\varepsilon-1)}{\alpha\varepsilon}\left(\frac{2}{3}+\frac{1}{3\varepsilon}\right)\right)-2\left(\frac{(\varepsilon-1)}{2\varepsilon+1}\right)^2+\frac{2}{3}\frac{(\varepsilon-1)}{2\varepsilon+1}-\frac{2\sigma}{RP_E}\right) \quad (3)$$

Figure 3 shows the time dependence of the pore size and the rate of expansion in water ($\sigma = 0.073$ N/m [33]) and in the transformer oil ($\sigma = 0.03$ N/m [34]), respectively. Accepted values of electrostrictive pressure $P_E$ correspond to the electric field in the fluid are close to the experimental conditions [20].

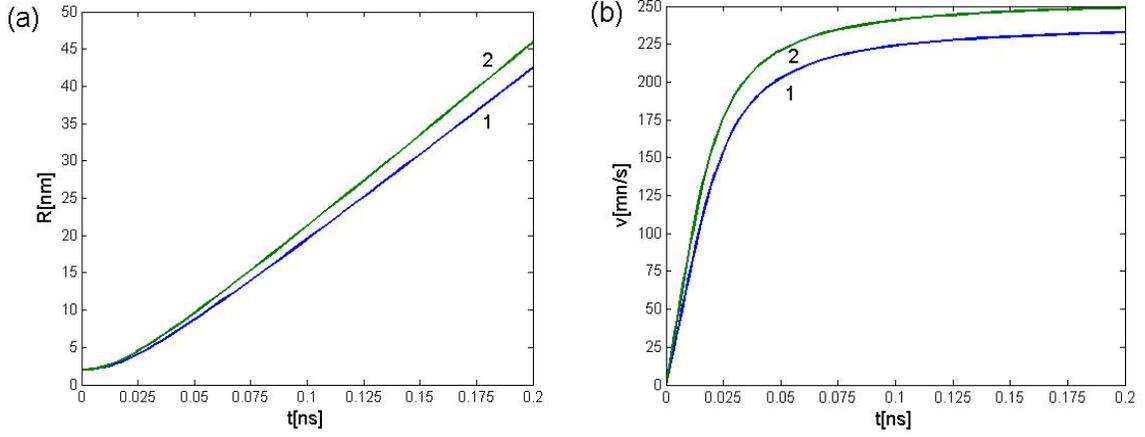

**Fig. 3**. Time dependencies of the pore size (a) and the speed of its expansion (b); 1 – in water ($P_E$=-250 MPa), 2 – in transformer oil ($P_E$=-300 MPa).

It can be seen that the rate of expansion of the pores reaches hundreds of meters per second, which on one hand they are almost by two orders of magnitude greater than the velocity of the fluid near the electrode [19], but on the other they are much smaller than the velocity of sound and reach the size of about 10 nanometers within a few hundreds of nanoseconds. Thus, for accepted values of "external" field $E_0$, the potential difference at the poles of pores reaches or exceeds the ionization potential, at $R \approx 5$ and at 14 nm, for water and oil, respectively.

**2. "Cavitation" breakdown in the liquid.**
As already noted, the negative electrostrictive pressure in the fluid, caused by the ponderomotive forces, induces the fluid influx, which not only compensates the electrostrictive stretching tension, but also leads to the appearance of the cavitation nanopores [18-24]. As can be seen from Figure 1, there are regions in the vicinity of the pores where the absolute values of the negative pressure are maximum, and therefore the cavitation development is most probable. Figure 4 shows the distribution of the radial forces $F_r = -\partial P_p / \partial r$ acting on the fluid, referred to $|P_E|/R$. It is seen that in the vicinity of the pole the forces are directed from the pore (fluid moves from the pore), but in the equatorial region - to the pore. Note, that for a pore in water the maximum force is located at a certain distance from the pole. Therefore, it can be expected that the dominant process in the equatorial region is a fluid influx to the pore, but in the polar region - the appearance of new cavitation nanopores. Such mechanism of a secondary cavitation ruptures generation should lead to the appearance of chains of nanopores, aligned along the electric field lines, in which the breakdown may be developed. Furthermore, this mechanism is an explosive, because the rate of cavitation ruptures generation depends strongly on the absolute value of the electrostrictive negative pressure $\Gamma \propto |P_E|^3 \exp(-A(\sigma,T)/|P_E|)$ [m$^{-3}$s$^{-1}$], [27], and does not depend on the rate of expansion and merging of nanopores.

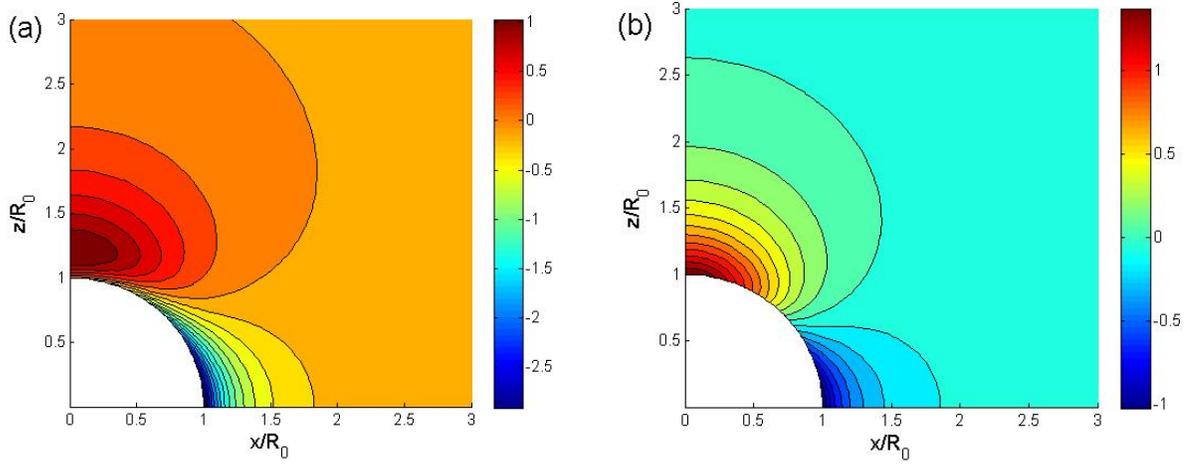

**Fig. 4.** Distribution of the radial forces $\widetilde{F}_r = -(\partial P_p/\partial r)/(|P_E|/R)$ in the vicinity of the nanopores. (a) is in water, (b) – in transformer oil.

## 3. Initiation of cavitation and nanosecond breakdown in oil on water microdroplets

Suppose that the microdroplets of water are located in the insulating transformer oil. In this case, the pressure associated with the electrostrictive ponderomotive forces is described by formula (2) wherein, $\varepsilon = 2.3$ and $\varepsilon_{in} = 80$. The distribution of electrostrictive negative pressure (in units $|P_E|$) around the microdroplets of water is shown in Fig. 5A. It is seen that in the vicinity of the water microdroplets pole the negative pressure is from seven to eight times greater than the pressure in the oil. And, in accordance with (2), the region of the elevated negative pressure grows with increasing of the radius of the droplet.

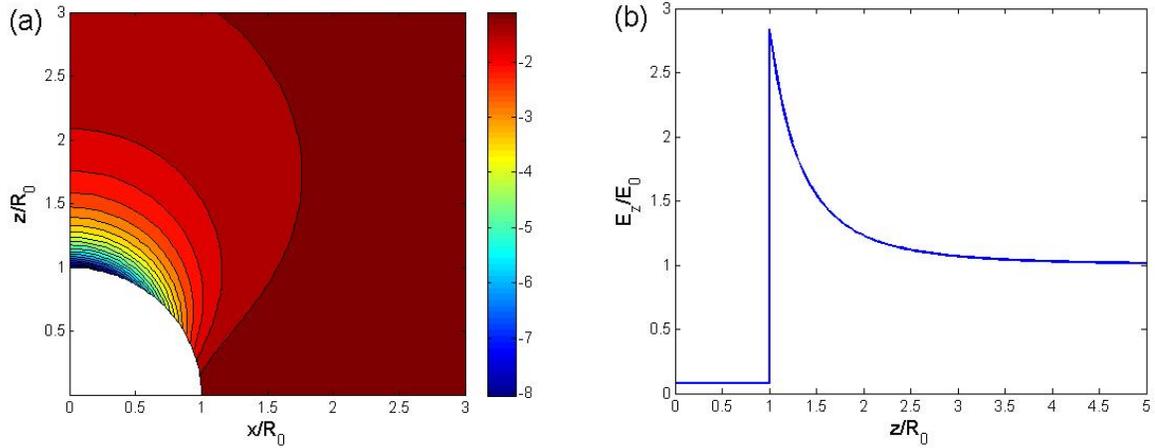

**Fig. 5.** (a) is the distribution of the electrostrictive pressure in transformer oil in the vicinity of water droplets (normalized to $|P_E|$); (b) is the corresponding electric field along the z axis, normalized to $E_0$.

The corresponding distribution of the electric field along the line passing through the droplet poles (along the z axis) is shown in Fig. 5B. It is seen that in the vicinity of the poles of water droplets, the electric field in the oil by more than twice exceeds the unperturbed field. That is, if the conditions for the initiation of the cavitation development and breakdown are not met in pure oil, these conditions may occur with large excess in the polar regions of water microdroplets. Thus, when the electric field is turned on quickly, water microdroplets may initiate breakdown in the transformer oil, i.e. determine its dielectric strength at essentially prebreakdown values of the electric field.

**Conclusions**

It is shown that the cavitation ruptures (nanopores) initiated by the electrostrictive ponderomotive forces in a liquid dielectric tend to be stretched in the direction of the electric field. A breakdown mechanism, associated with the generation of secondary cavitation ruptures in the vicinity of the poles of the original micropores, is proposed. It is shown that a local cavitation and electrical breakdown may develop in insulating oil in the vicinity of water microdroplets at substantially subcritical values of the electric field.